\documentstyle[preprint,prb,aps]{revtex}
\tightenlines

\begin{document}
\draft
\title{Strong enhancement of superconductivity \\ 
in a nanosized Pb bridge} 
\author{V. R. Misko\cite{A1}, V. M. Fomin\cite{A2}, J. T. Devreese\cite{A3},}
\address{Theoretische Fysica van de Vaste Stof,\\
Universiteit Antwerpen (U.I.A.),\\
Universiteitsplein 1, B-2610 Antwerpen, Belgium}
\date{\today}
\maketitle

\begin{abstract}
In recent experiments with a superconducting nanosized Pb bridge 
formed between a scanning tunneling microscope tip and a substrate, 
superconductivity has been detected at magnetic fields, which are 
few times larger than the third (surface) critical field. 
We describe the observed phenomenon on the basis of a numerical 
solution of the Ginzburg-Landau equations in a model structure consisting 
of six conoids. 
The spatial distribution of the superconducting phase is shown to be 
strongly inhomogeneous, with concentration of the superconducting phase 
near the narrowest part (the ``neck'') of the bridge. 
We show that suppression of superconductivity in the bridge by applied 
magnetic field or by temperature first occurs near the 
bases and then in the neck region, what leads to a continuous 
superconducting-to-normal resistive transition. 
A position of the transition midpoint depends on temperature and, 
typically, is by one order of magnitude higher than the second critical 
field $H_{c2}$. 
We find that the vortex states can be realized in the bridge at low 
temperatures $T/T_{c} \leq 0.6$. 
The vortex states lead to a fine structure of the 
superconducting-to-normal resistive transition. 
We also analyze vortex states in the bridge, which are characterized 
by a varying vorticity as a function of the bridge's height. 
\end{abstract}


\section{Introduction}

In a magnetic field parallel to the surface of a type II superconductor, the
nucleation of superconductivity \cite{sgst,gennes,tinkhis} near the surface
is known as ``surface superconductivity''. It is characterized by a third
critical field $H_{c3}$ which is higher than the second critical field $%
H_{c2}$ of the superconductor. E.g., for a plane superconductor-insulator
interface, the value of the third critical field is given by $%
H_{c3}=1.695H_{c2}$ \cite{sgst,gennes,tinkhis}. For magnetic fields between $%
H_{c2}$ and $H_{c3}$, a superconducting sheath appears near the surface. Its
thickness is of the order of the temperature-dependent coherence length $\xi
(T)$.

It is clear that the effect of surface superconductivity becomes
increasingly important with decreasing dimensions of the samples, in
particular when the volume of the near-surface layer becomes comparable to
the total volume of the sample. In this case one might expect that the
magnetic-field behavior of the mesoscopic superconductor is determined by
the third critical magnetic field $H_{c3}.$

However, not only the volume, but also the sample geometry plays a crucial
role in mesoscopic superconductors. The presence of several plane surfaces or
of a curved surface with a small radius of curvature (substantially smaller
than $\xi (T)$), enhances the superconductivity. As a result, critical
fields in mesoscopic structures even {\em exceed} $H_{c3}$, as will be shown
below theoretically and experimentally.

The nucleation of superconductivity at magnetic fields above $H_{c3}$ was
first studied in a wedge \cite{hml,vg}. Technological progress in the last
two decades enabled the manufacturing of mesoscopic superconducting
structures with sharp corners, and resulted in a renewed interest in the
problem of superconductivity in a wedge. For instance, recent investigations
on superconductivity in a wedge with a small angle $\alpha ,$ using a
variational approach \cite{bdfm,devr} or the adiabatic approximation \cite%
{devr,kfdm}, revealed that strongly localized distributions of the order
parameter dominate at $0<\alpha /\pi <0.1581$. Also, states of the
superconducting phase with integer numbers ($1$, $2$, $\ldots $) of confined
circulating superconducting currents have been found \cite{kfdm,devr} in
wedges with a sufficiently small angle ($\alpha /\pi \ll 1$). 
Within the framework of numerical solution of the GL equations in a wedge \cite{sp99}, 
the 
critical magnetic field for $\alpha=\pi/2$ was found to be approximately 
equal to $1.96H_{c2}$. 

Furthermore, strong enhancement of the superconductivity has been
demonstrated experimentally in Al mesoscopic squares and square loops, by
measuring the temperature dependence of the midpoint of the
normal-to-superconducting resistive transition \cite{mgs95}. The phase
boundaries obtained in this experiment lie in a region of substantially
higher magnetic fields than the phase boundary for bulk Al. The numerical
solution of the Ginzburg-Landau (GL) equations revealed \cite{wessc,weprb}
spatially inhomogeneous distributions of the order parameter in a mesoscopic
square loop, and the resulting superconducting phase boundary in a square
loop with leads \cite{weprb} turns out to be in agreement with the
experimentally observed phase boundary \cite{mgs95}. Also, the influence of
imperfections on the phase boundary of a superconducting mesoscopic double
loop has been analyzed \cite{devr,fdbm}. Recently, mesoscopic squares of
high-T$_{c}$ superconductors \cite{devr,wesscsd} were also studied.

Some theoretical work is moreover devoted to the superconducting properties
of mesoscopic disks and rings. Magnetic response of small superconducting
disks has been investigated \cite{dspg}, and first or second order
normal-to-superconducting transitions are found depending on the radius. 
A smooth transition from a multivortex superconducting state to a giant 
vortex state with increasing both the disk thickness and the magnetic field 
has been found for thin disks \cite{sp98}. 
The flux penetration and expulsion in thin superconducting disks have been
analyzed \cite{sp}. In Ref.~\onlinecite{palacios}, the saddle points or
energy barriers have been obtained, which are responsible for some
metastabilities observed in mesoscopic superconducting disks. On the basis
of the linearized GL equations, dimensional crossover in a mesoscopic
superconducting loop of finite width has been studied \cite{blvm}. It has
been shown that a dimensional transition occurs if the film thickness is of
order $\xi (T),$ similar to the 2D-3D transition for thin films in a
parallel magnetic field. Vortex states in superconducting rings with
out-of-center location of the opening have been recently discussed \cite{bps}. 

Superconductivity near the surface of a superconductor can still be enhanced
if it is surrounded by a medium with a higher transition temperature than
the one of the sample. The vortex structure of thin mesoscopic disks with
this type of enhanced surface superconductivity has been considered in Ref.~%
\onlinecite{yp}. The magnetic field versus temperature phase diagram was
obtained, and the regions of existence of the multivortex state and of the
giant vortex state were found. In Ref.~\onlinecite{monind} the enhancement
of the superconductivity near the surface of a mesoscopic superconductor was
studied, due to both the surrounding medium and the curved surface of the
sample.

All the above-mentioned superconducting structures can be described by 2D
(wedges, thin disks, squares and square loops) or quasi-3D (if the thickness
of a sample is effectively taken into account), or even 1D (double loop in a
network approach) GL equations. Another kind of mesoscopic superconducting
systems, which radically differs from the above structures, is a cone (or,
more generally, a conoid). Superconducting properties of a cone cannot be
described by 2D GL equations with an ``effective thickness'' of the sample.
Instead, the GL equations in a cone have to be solved in a general 3D form.

Our interest in the problem of superconductivity for a conic structure was
induced by recent experiments \cite{poza,sbbgv,urbva} with a nanosized Pb
bridge, which is formed between a substrate and the tip of a scanning
tunneling microscope. The shape of the bridge is close to a geometrical
figure consisting of two cones linked by their apexes. The diameter of the
bridge varies from a few nanometers to a few tens of nanometers. The
conductivity measurements showed a strong enhancement of the
superconductivity in the bridge. For magnetic fields as high as five times
their bulk critical value, the superconductivity survives in the bridge
while the leads become normal \cite{poza}.

In the present paper, we investigate a superconducting multi-conoid
structure using the GL theory. This study is relevant for the
above-mentioned experiments with a nanosized superconducting bridge. The
three-dimensional GL equations are solved self-consistently for the
superconducting order parameter and the magnetic field in the bridge. On the
basis of the obtained solutions, the superconducting-to-normal resistive
transition is calculated as a function of the applied magnetic field. 

The paper is organized as follows. The three-dimensional GL equations are
discussed in Sec.~II. In Sec.~III, we analyze the distributions of the order
parameter in the bridge as a function of the applied magnetic field and the
temperature. The vortex state is also investigated. The thermodynamical
stability of the solutions of the GL equations is studied in Sec.~IV for
different orbital quantum numbers $L$. Sec.~V is devoted to the 
description of the superconducting-to-normal resistive transition in the bridge. 
The 
possibility of the existence of superconducting states, characterized by
different values of $L$ along the bridge, is discussed in Sec.~VI.
Conclusions are given in Sec.~VII.

\section{GL equations in the bridge}

The GL equations for the order parameter $\psi $ and the vector potential $%
{\bf A}$ of a magnetic field ${\bf H=\nabla \times A}$ are \cite%
{gl50,gennes,tinkhis} 
\begin{eqnarray}
& &\frac{1}{2m}\left( -i\hbar {\bf \nabla }-\frac{2e}{c}{\bf A}\right) ^{2}\psi
+a\psi +b\left| \psi \right| ^{2}\psi =0,  \label{glepsi} \\
& &\Delta {\bf A} =\frac{4\pi ie\hbar }{mc}\left( \psi ^{\ast }{\bf \nabla }%
\psi -\psi {\bf \nabla }\psi ^{\ast }\right) +\frac{16\pi e^{2}}{mc^{2}}{\bf %
A}\left| \psi \right| ^{2}  \label{glea}
\end{eqnarray}%
with the boundary condition 
\begin{equation}
{\bf n}\cdot \left. \left( -i\hbar \nabla \psi -\frac{2e}{c}{\bf A}\psi
\right) \right| _{boundary}=0,  \label{bc}
\end{equation}%
where ${\bf n}$ is the unit vector normal to the boundary, $a$ and $b$ are
the GL parameters.

It is convenient to perform a transformation of Eqs.~(\ref{glepsi}), (\ref%
{glea}) to a dimensionless form. The new variables are defined as follows: 
\begin{equation}
\psi ^{\prime }=\frac{\psi }{\sqrt{\left| a\right| /b}},\ \ \ \ {\bf A}%
^{\prime }=\frac{2\pi \xi }{\Phi _{0}}{\bf A},\ \ \ \ x^{\prime }=\frac{x}{%
\xi },\ \ \ \ y^{\prime }=\frac{y}{\xi },  \label{tr}
\end{equation}%
where $\Phi _{0}=hc/2e$ is a flux quantum. The temperature-dependent
coherence length $\xi $ is given by 
\begin{equation}
\xi\equiv\xi (T)=\xi (0)\left( 1-\frac{T}{T_{c}}\right) ^{-1/2},  \label{xi} 
\end{equation}%
where $\xi (0)$ is the coherence length at zero temperature. The
temperature-dependent GL parameter $a$ has the form: 
\begin{equation}
a=a_{0}\left( 1-\frac{T}{T_{c}}\right) .  \label{a}
\end{equation}

From here on the primes will be omitted, and $\psi ,$ ${\bf A,}$ $x$ and $y$
will denote the scaled order parameter, vector potential and coordinates in
the units described in (\ref{tr}). As a result of the transformation (\ref%
{tr}), Eqs.~(\ref{glepsi}), (\ref{glea}) then become 
\begin{eqnarray}
& &\left( -i\nabla -{\bf A}\right) ^{2}\psi -\psi \left( 1-\left| \psi \right|
^{2}\right) =0,  \label{gledpsi} \\
& &\kappa ^{2}\Delta {\bf A} =\frac{i}{2}\left( \psi ^{\ast }{\bf \nabla }%
\psi -\psi {\bf \nabla }\psi ^{\ast }\right) +{\bf A}\left| \psi \right|
^{2}.  \label{gleda}
\end{eqnarray}%
Here $\kappa $ is the GL parameter defined as a ratio of the penetration
depth $\lambda (T)$ to the coherence length $\xi (T)$: 
\begin{equation}
\kappa =\frac{\lambda (T)}{\xi (T)},\ \ \ \ \lambda (T)=\lambda (0)\left( 1-%
\frac{T}{T_{c}}\right) ^{-1/2}.  \label{kappa}
\end{equation}%
The boundary condition (\ref{bc}) takes the form: 
\begin{equation}
{\bf n}\cdot \left. \left( -i\nabla -{\bf A}\right) \psi \right|
_{boundary}=0.  \label{bcd}
\end{equation}

As a result of solving Eqs.~(\ref{gledpsi}), (\ref{gleda}) with the boundary
conditions (\ref{bcd}), we shall obtain the spatial distributions of the
dimensionless order parameter $\psi $ and the dimensionless vector potential 
${\bf A}$. In order to return to dimensional variables and to reconstruct
their temperature dependence, we should afterwards recalculate them
according to the following rules.

\begin{enumerate}
\item Dimensional length $x_{dim}$ (similarly, $\rho_{dim}$, $z_{dim}$): 
\begin{equation}
x_{dim}=x\xi(0)\left(1-\frac{T}{T_{c}}\right)^{-1/2}.  \label{realx}
\end{equation}

\item Temperature-dependent superconducting order parameter: 
\begin{equation}
\psi (T)=\psi \sqrt{\left| a_{0}\right| /b}\left( 1-\frac{T}{T_{c}}\right)
^{1/2}.  \label{realpsi}
\end{equation}

\item Temperature-dependent vector potential: 
\begin{equation}
{\bf A}(T)={\bf A}\sqrt{2}\kappa H_{c}(0)\xi (0)\left(1-\frac{T}{T_{c}}\right)^{1/2},  
\label{reala}
\end{equation}%
where $H_{c}(T)$ is the thermodynamic critical magnetic field at 
temperature $T$, defined through the Helmholtz free energies $%
F_{n}(T)$ and $F_{s}(T)$ per unit volume in the normal and the
superconducting states \cite{tinkhis}, respectively: 
\begin{equation}
\frac{H_{c}^{2}(T)}{8\pi }=F_{n}(T)-F_{s}(T).  \label{hctd}
\end{equation}
\end{enumerate}

For the axial-symmetric problem, it is useful to introduce cylindrical
coordinates $(\rho ,z,\phi )$. The vector potential and the order parameter
can then be represented in the form 
\begin{equation}
{\bf A}(\rho ,z)={\bf e}_{\phi }A(\rho ,z)  \label{ac}
\end{equation}%
\begin{equation}
\psi (\rho ,z,\phi )=f(\rho ,z)e^{iL\phi },  \label{psiexp}
\end{equation}%
where $L$ is the orbital quantum number (which is a measure of the vorticity
of the superconducting state in the bridge).

The GL equations for the order parameter and for the vector potential
become: 
\begin{eqnarray}
&&\frac{1}{\rho }\frac{\partial f}{\partial \rho }+\frac{\partial ^{2}f}{%
\partial \rho ^{2}}+\frac{\partial ^{2}f}{\partial z^{2}}=f\left\{ \frac{%
L^{2}}{\rho ^{2}}-\frac{2AL}{\rho }+A^{2}-\left( 1-f^{2}\right) \right\} ,
\label{gledcpsi} \\
&&\kappa ^{2}\left\{ \frac{\partial }{\partial \rho }\left[ \frac{1}{\rho }%
\frac{\partial (\rho A)}{\partial \rho }\right] +\frac{\partial ^{2}A}{%
\partial z^{2}}\right\} =f\left\{ A-\frac{L}{\rho }\right\} .  \label{gledca}
\end{eqnarray}

It should be noted that Eqs.~(\ref{gledcpsi}), (\ref{gledca}) in the bridge
have to be solved in their general {\it three-dimensional} form. Due to the
complicated shape of the bridge, simplifications of Eqs.~(\ref{gledcpsi}), (%
\ref{gledca}) are not allowed, as, for example, in the case of a disk \cite%
{dspg}. Mesoscopic disks considered in Ref.~\onlinecite{dspg} were of a
finite width $d\sim \xi $. Because of this condition, the order parameter
was supposed to be uniform along the axis of the disk \cite{dspg} resulting
in the disappearance of the third term in the left-hand side of Eq.~(\ref%
{gledcpsi}). Another approximation, which led to a substantial
simplification of Eq.~(\ref{gledcpsi}) in the case of a mesoscopic disk \cite%
{dspg}, consisted in the substitution of a local value of the vector
potential $A$ obtained from Eq.~(\ref{gledca}) by its average $\langle
A\rangle $ over the thickness of the disk. Obviously, these approximations
cannot be applied for the problem of the mesoscopic bridge under
consideration, which consists of a few conoids, and we actually have to deal
with a three-dimensional problem.

At the conoidal surface with an angle $\alpha $, Eq.~(\ref{bcd}) results in
the following boundary condition for the function $f(\rho ,z)$: 
\begin{equation}
\left. \left( \frac{\partial f}{\partial \rho }\cos \,\alpha +\frac{\partial
f}{\partial z}\sin \,\alpha \right) \right| _{boundary}=0.  \label{bcdcpsi}
\end{equation}%
At the 
base, 
we have: 
\begin{equation}
\left. \frac{\partial f}{\partial z}\right| _{boundary}=0.  \label{bcdcpsib}
\end{equation}%
For the vector potential, the boundary condition is determined at infinity 
\begin{equation}
\left. A\right| _{\infty }={\bf e}_{\phi }H_{0}\rho /2,  \label{bcdca}
\end{equation}%
where $H_{0}$ is the applied magnetic field. 
Because of nonzero demagnetizing factor, the local magnetic field 
at the surface of the bridge differs from the applied magnetic field. 
However, our calculations show that already at a distance of about 
two maximal radii of the bridge, the effect of magnetic field 
distortion by the sample is negligible. 
Hence, for the purpose of numerical calculations, we choose 
a simulation region, which is large enough (see below) to assume 
that all changes of the magnetic field occur {\it inside} this region, 
while outside of it 
the magnetic field is 
uniform and equal to the applied magnetic field. 
Therefore, the 
boundary condition (\ref{bcdca}) is substituted by 
\begin{equation}
\left. A\right| _{bsr}={\bf e}_{\phi }H_{0}\rho /2,  \label{bcdcasr}
\end{equation}%
where ``$bsr$'' denotes the boundary of a simulation region.

As a realistic model of the bridge used in Ref.~\onlinecite{poza}, we choose
a geometrical figure restricted by six conoidal surfaces and two 
bases (Fig.~1). We place the bridge into the three-dimensional Cartesian frame of 
reference with the $z$-axis coinciding with the symmetry axis of the bridge.
The magnetic field is supposed to be applied along the $z$-axis. The point $%
z=0$ is chosen in the plane, where the bridge has its minimal diameter. This
narrowest cross-section of the bridge will be referred to as the {\it neck}.
The heights of the conoids and their diameters are indicated in Fig.~1. The
conoidal surfaces forming the boundaries of the bridge are then described by
the following equations: 
\begin{eqnarray}
z=\frac{2h_{1}}{b_{1}-b_{0}}\left( \sqrt{x^{2}+y^{2}}-\frac{b_{0}}{2}\right)
,\ &&\ \text{if }\ 0\leq z\leq h_{1},  \label{cones} \\
z=\frac{2h_{2}}{b_{2}-b_{1}}\left( \sqrt{x^{2}+y^{2}}-\frac{b_{1}}{2}\right)
+h_{1},\ \ &&\text{if }\ h_{1}<z\leq h_{1}+h_{2},  \nonumber \\
z=\frac{2h_{3}}{b_{3}-b_{2}}\left( \sqrt{x^{2}+y^{2}}-\frac{b_{2}}{2}\right)
+h_{1}+h_{2},\ \ &&\text{if }\ h_{1}+h_{2}<z\leq h_{1}+h_{2}+h_{3}. 
\nonumber
\end{eqnarray}

The boundary conditions (\ref{bcdcpsi}) are applied at the conoidal surfaces
(\ref{cones}) with $\alpha =\alpha _{1}$, $\alpha _{2}$, $\alpha _{3}$: 
\begin{eqnarray}
\alpha _{1} &=&\arctan \left( \frac{b_{1}-b_{0}}{2h_{1}}\right) , \\
\alpha _{2} &=&\arctan \left( \frac{b_{2}-b_{1}}{2h_{2}}\right) ,  \nonumber
\\
\alpha _{3} &=&\arctan \left( \frac{b_{3}-b_{2}}{2h_{3}}\right) .  \nonumber
\label{alphas}
\end{eqnarray}%
At the surface 
\begin{equation}
z=h_{1}+h_{2}+h_{3}\ \text{with}\ \ x^{2}+y^{2}\leq \left( \frac{b_{3}}{2}%
\right)^{2}, \label{base} 
\end{equation}%
the boundary condition (\ref{bcdcpsib}) is applied.

The part of the bridge situated in the area $z<0$ is described by equations
symmetric to Eqs.~(\ref{cones}) to (\ref{base}) with respect to the plane $%
z=0$.

The self-consistent numerical solutions of Eqs.~(\ref{gledcpsi}), (\ref%
{gledca}) with the boundary conditions Eqs.~(\ref{bcdcpsi}) to (\ref{bcdca})
in the bridge are obtained using the finite difference method for solving
partial differential equations \cite{smith}. As a simulation region in the
numerical calculations, we choose a cylinder with a radius which is four
times the maximal radius of the bridge, and with a height which is three
times the height of the bridge. This choice has been proven to be sufficient
for obtaining solutions which are independent of the size of the simulation
region. 
The relative accuracy of the obtained distributions of the order parameter 
is 10$^{-4}$. 
The spatial resolution of the computational method is about 0.02$\xi(0)$. 

\section{Solutions of the Ginzburg-Landau equations in the bridge:
distribution of the superconducting phase}

\subsection{Magnetic-field dependence of the distribution of the
superconducting phase in the bridge}

We first discuss some typical features of the distribution of the squared
amplitude $\left| \psi (x,y,z)\right| ^{2}$$\equiv \left| \psi \right| ^{2}$
of the order parameter in the bridge, which are represented in Fig.~2. 
All
the distributions of $\left| \psi \right| ^{2}$ throughout this paper are
shown as contour plots of the cross-section along the symmetry axis of the
bridge, i.e. along the plane $y=0$ (see Fig.~1). 
The solutions of the GL equations, which we discuss in this paper, are obtained 
for $\kappa=3.9$. 
Because $\kappa$ is the only parameter entering 
Eqs.~(\ref{gledcpsi}) and (\ref{gledca}), 
the obtained solutions are applicable for different values of the coherence length. 
In order to emphasize this general character of the obtained solutions, we express 
the sizes in the plots in units $\xi\equiv\xi (T)$ as defined by Eq.~(\ref{xi}). 
The sizes expressed in nanometers, also indicated in the plots, 
correspond to the parameters of the Pb bridge \cite{poza}, for which 
$\xi (0)$ is estimated to be 10 nm (see subsection III.C.). 
In Fig.~2(a), $\left| \psi
\right| ^{2}$ is plotted for the temperature $T/T_{c}=0.6$, for the applied
magnetic field $H_{0}=H_{c2}$, where $H_{c2}=\sqrt{2}\kappa H_{c}(T)$, 
and for the orbital quantum number $L=0$.
The superconducting phase is concentrated near the neck of the bridge. The
function $\left| \psi \right| ^{2}$ has the maximal possible value equal to
1 in the denotations (\ref{tr}) within a region, which symmetrically spreads
along the $z$-axis to the distances of about $\xi$ from the plane $z=0$. In 
this region, the superconductivity is strongly enhanced due to the small
lateral dimensions of the neck. There are no significant changes of $\left|
\psi \right| ^{2}$ as a function of magnetic field in this area (Fig.~2).
Away from the neck, at distances of about $\xi$ to 2$\xi$ from the plane $z=0$, $%
\left| \psi \right| ^{2}$ gradually decreases down to the value $\left| \psi
\right| ^{2}\sim 0.5$. Qualitatively, such a behavior of $\left| \psi
\right| ^{2}$ is typical for various temperatures and applied magnetic
fields.

An increasing applied magnetic field suppresses the superconductivity in the
bridge starting from its 
bases. 
For the applied magnetic field $H_{0}=1.6H_{c2}$ 
[Fig.~2(b)] and the same temperature and $L$ as in the
previous panel, ring-shaped areas with $\left| \psi \right| ^{2}\sim 0.4$
appear near the bases of the bridge (represented in the contour plot as 
small areas in the corners). For a higher applied magnetic field 
$H_{0}=2H_{c2}$, these areas still grow. It is worth noting that the area 
near the neck of the bridge does not noticeably change with applied magnetic
field.

For $H_{0}=4H_{c2}$ (all other parameters are the same in Fig.~2), the 
area near the neck, which is filled by the superconducting phase with the
maximal value of $\left| \psi \right| ^{2}=1$, is reduced approximately by a
factor of 0.5 as compared to the case $H_{0}=H_{c2}$ [cf. Fig.~2(a)]. The 
function $\left| \psi \right| ^{2}$ changes from its maximal value near $z=0$
down to $\left| \psi \right| ^{2}\sim 0.3$ near the bases. 

A further increase of the applied magnetic field leads to a stronger
suppression of the superconducting state in the bridge. For the applied
magnetic fields $H_{0}=5H_{c2}$ and $H_{0}=8H_{c2}$ [Figs.~2(e) and~2(f)], 
the function $\left| \psi \right| ^{2}$ decreases very fast towards the
bases of the bridge. At $H_{0}=12H_{c2}$ the bases of the bridge 
are connected to each other by an area with $\left| \psi \right| ^{2}$
ranging from 0.01 to 0.09, and some regions in the bridge are filled by the
normal state.

Finally, at $H_{0}=16H_{c2}$ the superconducting phase is concentrated
near the neck only. It is separated from the bases of the bridge by 
regions of the normal state [Fig.~2(h)]. We believe that such a pattern can be
experimentally detected by measuring the density of states in the tunneling
regime as a function of the applied magnetic field \cite{suderow}.

\subsection{The distribution of the superconducting phase in the bridge as a
function of temperature}

The distribution of the superconducting phase in the bridge is strongly
temperature-dependent. In this subsection, the temperature dependence of the
squared amplitude $\left| \psi \right| ^{2}$ of the order parameter is
studied. 

We start with the distribution of $\left| \psi \right| ^{2}$ for $T/T_{c}=0.6
$ [Fig.~3(a)]. This distribution is characterized by decreasing values of $%
\left| \psi \right| ^{2}$ with increasing $\left| z\right| $ near the area
of the neck, and with increasing $\left| x\right| $ near the bases of 
the bridge. At a higher temperature [$T/T_{c}=0.9$, Fig.~3(b)], the
distribution of $\left| \psi \right| ^{2}$ is substantially modified as
compared to the previous case. It becomes uniform along the $z$-axis near
the bases. 
The typical butterfly-like pattern remains near the area of
the neck only. This central area is filled by the superconducting phase with
the maximal value of $\left| \psi \right| ^{2}=1$. 
With further increasing temperature up to $T/T_{c}=0.99$ [Fig.~3(c)], the area 
near the 
neck of the bridge becomes separated from the bases by wide regions, which 
are characterized by a small value of $\left| \psi \right| ^{2}$ $<0.1$. 
For higher temperatures, the function $\left| \psi \right| ^{2}$ 
remains nonzero only near the area of the neck and vanishes very fast 
versus $z$. 

With increasing temperature, the superconducting phase in the nanosized
superconducting bridge is thus suppressed by two mechanisms. Along with a
uniform reduction of the maximal value of $\left| \psi \right| ^{2}$,
described by a simple dependence $1-T/T_{c}$ following from Eq.~(\ref{realpsi}), 
also a spatial redistribution of $\left| \psi \right| ^{2}$ 
takes place in the bridge when the temperature increases. This
redistribution is characterized by a concentration of the superconducting
phase near the neck of the bridge accompanied by a relative suppression of $%
\left| \psi \right| ^{2}$ away from this central area.

\subsection{The vortex state in the bridge}

Most of the bulk superconducting metals are typically type I
superconductors, characterized by a Ginzburg-Landau parameter $\kappa \sim
0.03$ in Al to $\kappa \sim 0.48$ in Pb \cite{poole}. However, in
mesoscopic systems, metals can become type II superconductors
(like most superconducting metal alloys and compounds \cite{poole}). 
This is due
to a substantial reduction of the effective coherence length in mesoscopic
systems, where the size of the sample plays the role of the electron mean
free path in bulk systems. As a result, the parameter $\kappa $ becomes
larger than $1/\sqrt{2}$ in mesoscopic metallic structures. For example, in
mesoscopic square loops of Al studied in the experiment \cite{mgs95}, the
parameter $\kappa $ was found to be $\sim 1$. This experimentally obtained
value of $\kappa $ was used in the calculations of the phase boundaries of a
mesoscopic square loop \cite{wessc,weprb}.

For the Pb bridge used in Ref.~\onlinecite{poza}, the effective coherence
length is determined by the bridge diameter, which varies from a few
nanometers in the vicinity of the neck to about 23 nm near the bases. 
In our calculations, we estimated the value of the effective coherence length 
to be 10 nm \cite{uncert}. According to Ref.~\onlinecite{poole}, the
penetration depth at zero temperature for Pb is $\lambda(0)=39$ nm, and consequently 
$\kappa = 3.9$. Hence, the material of the bridge is effectively a type 
II superconductor. Therefore, a vortex state can be realized in the bridge. 

We have calculated the distributions of the superconducting phase in the
bridge for different orbital quantum numbers $L$. In Fig.~4, such
distributions are shown for the temperature $T/T_{c}=0.6$ and 
for the applied magnetic field $H_{0}=5H_{c2}$. 
The distributions of
the superconducting phase are presented for $L=0$ to $L=3$ in Fig.~4. The
function $\left| \psi \right| ^{2}$ is strongly modified by the presence of
the vortex in the bridge, as compared to the case without a vortex ($L=0$).
For $L>0$, the distributions have the shape of a sheath near the conic surface. 
The thickness of the sheath varies in different parts
of the bridge (see Fig.~4). In addition, the function $\left| \psi \right|
^{2}$ is spatially inhomogeneous in the sheath. The maximal value of $\left|
\psi \right| ^{2}$ is reached in the vicinity of the bridge surface in the
area near the neck. The function $\left| \psi \right| ^{2}$ decreases
towards the axis of the bridge $z=0$. These changes are shown in Fig.~4. In
order to determine, which of these states is realized in the bridge, we now
turn to the problem of the thermodynamical stability of solutions of the GL
equations for different values of $L$.

\section{The thermodynamical stability of the solutions of the
Ginzburg-Landau equations in the bridge}

The free energy per unit volume of a superconductor in the magnetic field is %
\cite{gl50,parks}: 
\begin{equation}
F_{s}=F_{n}+a\left| \psi \right| ^{2}+\frac{b}{2}\left| \psi \right| ^{4}+%
\frac{H^{2}}{8\pi }+\frac{1}{2m}\left| -i\hbar {\bf \nabla }\psi -\frac{2e}{c%
}{\bf A}\psi \right| ^{2}-\frac{H_{0}^{2}}{8\pi },  \label{fs}
\end{equation}%
where $F_{n}$ is the free energy per unit volume of the normal (non-superconducting) phase. 
The superconducting state is stable when the difference $F_{s}-F_{n}$ has a 
minimum. 

It can be shown \cite{abr} that the difference $F_{s}-F_{n}$ can be
expressed as follows: 
\begin{equation}
F_{s}-F_{n}=\frac{H_{c}^{2}(0)}{4\pi }\left[ \left( H-H_{0}\right) ^{2}-%
\frac{1}{2}\left| \psi \right| ^{4}\right] .  \label{fsnpsi4}
\end{equation}%
In the dimensionless form, the free energy Eq.~(\ref{fsnpsi4}) calculated
over the bridge of volume $V_{b}$ then takes the form: 
\begin{equation}
\int\limits_{V_{b}}\left( F_{s}-F_{n}\right) dxdydz=\int\limits_{V_{b}}\left[
\left( H-H_{0}\right) ^{2}-\frac{1}{2}\left| \psi \right| ^{4}\right] dxdydz.
\label{fsnpsi4d}
\end{equation}

Before calculating the free energy of the superconducting bridge, let us
estimate contributions of two terms in the right-hand side of Eq.~(\ref%
{fsnpsi4d}) to the integral in the left-hand side of Eq.~(\ref{fsnpsi4d}).
The first term describes the contribution to the free energy due to changes
in the magnetic field in the bridge as compared to the applied magnetic
field. These changes occur over distances of the order of the penetration
depth $\lambda $ of the magnetic field. The temperature dependent values of $%
\lambda (T)$ calculated using Eq.~(\ref{kappa}) are listed in Table~1 for
various temperatures [recall that $\lambda(0)=39$ nm\cite{poole}]. 

\begin{center}
Table~1. The magnetic field penetration depth in Pb for various
temperatures. \\[0pt]
\bigskip 
\begin{tabular}{|c|c|c|c|c|}
\hline
$T/T_{c}$ & 0.6 & 0.9 & 0.99 & 0.998 \\ \hline
$\lambda(T)$, nm & 61.7 & 123.4 & 390.0 & 782.5 \\ \hline
\end{tabular}
\end{center}

According to Table~1, the values of $\lambda (T)$ are larger than the mean
diameter of the bridge (10 nm). Therefore, in the case $L=0$ we can
approximately disregard the changes of the magnetic field in the bridge and
omit the first term $(H-H_{0})^{2}$ in the right-hand side of Eq.~(\ref%
{fsnpsi4d}). For $L>0$ this also seems an acceptable approximation because $%
\kappa $ is large and the lateral dimensions of the bridge are small.
Indeed, for large values of $\kappa $ changes in the magnetic field occur
within a relatively large region with sizes of order $2\lambda (T),$  in
contrast to the sharp changes of the order parameter, which occur within a
relatively small region with sizes of the order of $2\xi (T)$. The magnetic
field changes in the sample because the system captures $L$ flux quanta $%
L\Phi _{0}$. The amplitude of the changes in the magnetic field is small if
they occur within a large region. Recall that the bridge diameter is much
smaller than the size of the region of magnetic field changes [that is $\sim
2\lambda (T)$]. In summary, the contribution $\int_{V_{b}}(\Delta
H)^{2}dxdydz$ in Eq.~(\ref{fsnpsi4d}) decreases also for $L>0$ with
increasing $\kappa $.

We have calculated the free energy of the superconducting states in the
bridge as a function of the applied magnetic field in a wide range of
applied magnetic fields from $0$ to $30H_{c2}$. 
The calculations are
performed for three values of the temperature: $T/T_{c}=0.4$, $T/T_{c}=0.6$
and $T/T_{c}=0.9$, and for various values of the orbital quantum number $L$.

In Fig.~5, the free energy of the bridge [measured in units of $%
H_{c}^{2}(0)/4\pi $] is plotted as a function of the applied magnetic field
for the temperature $T/T_{c}=0.4$. The states characterized by $L=0$ to $L=4$
are shown. For the range of magnetic fields 0 to $5H_{c2}$, the state $L=0$ 
minimizes the free energy. In this magnetic field range  the values of $%
\left| F_{s}-F_{n}\right| $ are about three times larger for the state $L=0$
than for the state $L=1$. For applied magnetic fields higher than $5H_{c2}$, 
the states with $L=1$ to $L=4$ can be realized. But the values of $\left|
F_{s}-F_{n}\right| $ for these $L$ are far less sensitive to the value of $L$
than they are in the interval of magnetic fields from 0 to $5H_{c2}$. 
In Fig.~6, the free energy of the bridge is shown as a function of the applied 
magnetic field at the temperature $T/T_{c}=0.6$. The range of the applied 
magnetic fields, for which the state $L=0$ is realized, becomes more than 
two times wider as compared to $T/T_{c}=0.4$ (cf. Fig.~5): the upper value 
of this range increases until about $10.5H_{c2}$. 
In Fig.~7, the free
energy of the bridge is plotted as a function of the applied magnetic field
for $T/T_{c}=0.9$. In this case, the state $L=0$ is realized for all the
applied magnetic fields in the range 0 to $30H_{c2}$. 
Therefore, penetration of a vortex into the bridge is only possible 
for sufficiently small temperatures. 

\section{The superconducting-to-normal resistive transition for the bridge} 

In Ref.~\onlinecite{poza}, a narrow cylinder was used as an oversimplified
model of the bridge. We have taken into account a more complicated shape of the
bridge (compare Fig.~1 in Ref.~\onlinecite{poza} to Fig.~1 in the present
work). Our self-consistent numerical solutions of the three-dimensional GL
equations have shown that the distribution of the superconducting order
parameter is strongly inhomogeneous in the bridge. In the superconducting
state, the bases of the bridge are connected to each other by a 
continuous superconducting sheath. The width of the sheath is different in
various parts of the bridge. Also, the value of $\left| \psi \right| ^{2}$
is a position-dependent. With increasing magnetic field (see Fig.~2) or
increasing temperature (see Fig.~3), the superconducting properties 
decrease continuously. Analyzing the distributions of the superconducting
phase, we can expect that the superconducting-to-normal transition as a
function of applied magnetic field or temperature has the shape of a
smoothed step function. For example, consider the distributions of the
superconducting phase shown in Fig.~2. 
The resistance is expected to be minimal for 
the applied magnetic field $H_{0}=H_{c2}$ [see Fig.~2(a)]. With increasing 
applied magnetic field, the value of $\left| \psi \right| ^{2}$ reduces near
the bases, and therefore the resistance of the bridge increases. 
For the applied magnetic field $H_{0}=16H_{c2}$ [Fig.~2(h)], the 
resistance of the bridge is close to that of the normal metal.
After these intuitive considerations, we now turn to the calculations of the 
resistance of the bridge as a function 
of the applied magnetic field. 

To calculate the resistance of the bridge, we divide the bridge into thin
disks, each with thickness $\delta$ substantially smaller than the coherence length 
$\xi (T)$. Therefore, the function $\left| \psi \right| ^{2}$ only changes in
the $xy$-plane of the disk. Each disk is subdivided into thin rings with 
width $\delta$. 
As the smallest partition element (``cell'') of the bridge, we choose 
a ring sector with length $\delta$. 
Because of the condition $\delta\ll\xi(T)$ (recall that $\xi(T)$ 
is chosen as a unit of length), 
the volume of the cell is independent of the radius of the ring and is approximately 
equal to $\delta^{3}$. 
The resistance of each cell $R_{cell}$ 
is equal to zero, if the cell is in a superconducting state, and it is equal to 
the normal resistance $R_{cell,n}$, if the cell is in a normal state. 
For the numerical calculations, the resistance of the cell (measured in $R_{cell,n}$) 
is represented in the form 
\begin{equation} 
R_{cell}=
\left\{
\begin{array}{ccc} 
0, & \mbox{if} & \mid\psi(x,y,z)\mid^{2} > \epsilon \\ 
1, & \mbox{if} & \mid\psi(x,y,z)\mid^{2} \leq \epsilon 
\end{array}
\right., 
\label{rcell} 
\end{equation} 
where $\epsilon \ll 1$. 
Equation (\ref{rcell}) can be approximated by the analytical 
expression 
\begin{equation} 
R_{cell}\approx\frac{\epsilon}{\mid\psi(x,y,z)\mid^{2} + \epsilon}, 
\label{rcellan} 
\end{equation} 
which asymptotically coincides with Eq.~(\ref{rcell}) when $\epsilon \to 0$. 
In the calculations executed using Eq.~(\ref{rcellan}), we choose $\epsilon = 0.001$ 
[reduction of the value $\epsilon$ down to 0.0001 
does not lead to any appreciable changes of the curve $R=R(H_{0})$]. 
Because of cylindrical symmetry, 
the resistance of the ring with radius $r$ is $\delta R_{cell}/2\pi r$. 
The total resistance of the bridge is obtained based on the formulas 
for parallel and series resistances. 

In Fig.~8(a), the resistive superconducting-to-normal transition is shown as a
function of the applied magnetic field for the temperature $T/T_{c}=0.4$. For
the fields ranging from 0 to approximately $5H_{c2}$, the bridge is 
superconducting and has zero resistance. The resistive
superconducting-to-normal transition is characterized by a range of fast
changes in the interval $5H_{c2}$ to $15H_{c2}$, followed by a range of 
slow changes for higher fields. The above-mentioned fast changes are
explained by a fast reduction of $\left| \psi \right| ^{2}$ in the regions
near the bases of the bridge with increasing applied magnetic field. In 
contrast, the middle part of the distribution $\left| \psi \right| ^{2}$
varies slowly versus applied magnetic field, leading to the above-stated
slow changes of the resistive superconducting-to-normal transition.

In order to compare the resistive transitions calculated for different 
temperatures, we introduce an additional scale (indicated on top of each 
panel) for the applied magnetic field $H_{0}$ measured in $H_{c2}$ at zero 
temperature. 
Figs.~8(b) and 8(c) show that 
for higher temperatures, $T/T_{c}=0.6$ and $T/T_{c}=0.9$, the resistive
superconducting-to-normal transition is shifted to lower applied magnetic
fields as compared to the case $T/T_{c}=0.4$. 

Taking into account the states with orbital quantum numbers $L>0,$ steps 
appear in the resistance as a function of the magnetic field, as shown in 
Fig.~8b. It is worth noting that the resistive superconducting-to-normal
transition for the states with $L>0$ is close to that for $L=0$. The
contribution of the states with $L>0$ to the resistive
superconducting-to-normal transition in the mesoscopic bridge is, however,
relatively small as compared to the state with $L=0$.

\bigskip

In summary, the resistive superconducting-to-normal transition as a function
of applied magnetic field is characterized by a fast change in some range of
relatively low applied magnetic fields and by a slow change for higher
fields. This behavior is explained by an inhomogeneous distribution of the
superconducting phase in the bridge. The fine structure appears in the
resistive superconducting-to-normal transition as a 
function of applied magnetic field, when the states with $L>0$ are taken
into account.

\section{The superconducting states characterized by different $L$ along the
bridge}

As discussed above, the diameter of the bridge varies along the $z$-direction 
from a few nanometers in the neck ($z=0$) until about 23 nanometers near the 
bases. 
This suggest that, given the applied magnetic
field and the temperature, superconducting states characterized by various $L$ 
are realized in different parts of the bridge. 
In this section, we analyze
the possibility of existence of such vortex states. For this purpose, we
calculate the free energies of the states with various vorticities for disks
of different radii. The radii of the disks are chosen to coincide with those
of cross-sections of the bridge in the $xy$-planes at different values of $z$. 
The results of the free energy calculations are shown in Fig.~9. For
sufficiently large disks [Fig.~9(a), $D_{disk}=1.5\xi$ that corresponds to 24 nm 
in the case of the Pb bridge (Ref.~\onlinecite{poza}) at $T/T_{c}=0.6$], the states 
characterized by different $L$ are possible for various applied magnetic
fields. The transition $L=0\rightarrow L=1$ takes place at about 
$H_{0}=3H_{c2}$ for the temperature $T/T_{c}=0.6$ [Fig.~9(a)]. For disks with 
smaller radii, this transition occurs at higher values of the applied
magnetic field [Figs.~9(b) to 9(d)]. For disks with very small radii, only the
state $L=0$ is realized in the considered range of applied magnetic fields
[Fig.~9(e)].

Using the obtained plots (Fig.~9), we can find the radii of the disks at
which a transition $L\rightarrow L+1$ occurs for different values of the
applied magnetic field. For $H_{0}=5H_{c2}$ the transition $L=0\rightarrow 
L=1$ turns out to take place in a disk with diameter $D_{disk}=1.125\xi$ 
[18 nm for the Pb bridge (Ref.~\onlinecite{poza}) at $T/T_{c}=0.6$], 
shown in Fig.~9(c). For the area between the neck of the bridge and the disk, which
is shown in Fig.~9(c), the state $L=0$ is realized. For the areas between the
bases of the bridge and the disk [Fig.~9(c)], the state $L=1$ is 
energetically favored. Thus, there are three competing states: the state
with $L=0$ all over the bridge [Fig.~10(a)], that with $L=1$ all over the
bridge [Fig.~10(b)], and also the vortex state characterized by a {\it varying
vorticity} $L=\{0,1\}$ in the direction of the bridge's axis [Fig.~10(c)]. We have 
calculated the free energies for these states. The calculations show that
the state $L=\{0,1\}$ with varying vorticity has the smallest free energy.
The value of the free energy for this state is about 50 per cent lower than
for the state $L=1$. However, the free energy for the state $L=\{0,1\}$ is
only 2 per cent lower than that for the state $L=0$.

Hence, in the bridge under consideration the state $L=0$ is a reasonable
approximation for a wide range of applied magnetic fields. Comparison in
Fig. 8c between the resistance calculated for thermodynamically equilibrium
states involving different numbers $L$, on the one hand, and the resistance
calculated only for the state with $L=0$, on the other hand, confirms this 
conclusion. But to describe the fine structure of the
superconducting-to-normal resistive transition [Fig.~8(b)], 
the states with $L>0$ have to be taken into 
consideration.

\section{Conclusions}

We have solved the {\it three-dimensional} Ginzburg-Landau (GL) equations
in a mesoscopic superconducting multi-conoidal structure with cylindrical
symmetry. The calculations have been performed for a geometrical body, which
is restricted by six conoidal surfaces and two base plane surfaces. This 
body is a model of a Pb nanosized superconducting bridge between a
scanning tunneling microscope tip and a substrate \cite{poza}. 
Our calculations show that the 
distribution of the superconducting phase in the bridge is strongly
inhomogeneous.

Superconductivity in the bridge is suppressed by increasing the applied
magnetic field. This suppression occurs first near the bases of the 
bridge. In a region near the neck of the bridge, superconductivity survives
to much higher magnetic fields than the surface critical field $H_{c3}$. At
high magnetic fields this neck region is separated from the bases by 
regions of the normal phase.

We have also studied the suppression of superconductivity in the bridge by
increasing temperature. The maximal value of $\left| \psi \right| ^{2}$
decreases along with a redistribution of $\left| \psi \right| ^{2}$ in the
bridge. This redistribution is characterized by a concentration of the
superconducting phase near the neck of the bridge with a relative reduction
of $\left| \psi \right| ^{2}$ away from this central region.

We have analyzed the superconducting states in the bridge from the point of
view of the thermodynamical stability. The free energy of the
superconducting states has been calculated as a function of the applied
magnetic field. For a wide range of applied magnetic fields, the state $L=0$
is realized in the bridge. This range widens with increasing temperature.

The superconducting-to-normal resistive transition is studied in detail as a
function of the applied magnetic field. We have shown that a switching from
the superconducting to the normal regime occurs in a wide interval of
applied magnetic fields of the order of $10H_{c2}$ and depends on 
temperature. For example, for $T/T_{c}=0.6$, the resistive transition starts
at about $5H_{c2}$ and reaches the midpoint at about $15H_{c2}$. 
This agrees with the experimental data according to which superconductivity in 
the bridge has been detected at $5H_{c2}$ \cite{poza}. It is shown, that 
the main contribution to the superconducting-to-normal resistive transition
is determined by the state with orbital angular momentum $L=0$. When states
with higher $L$ are taken into account, a fine structure appears in the
transition, which reflects the effect of the fluxoid quantization in the
bridge.

In the bridge, the vortex states can exist, which are
characterized by various $L$ along the height of the bridge. Namely, it is
shown that, at applied magnetic fields higher than $5H_{c2}$, the vortex 
state with a varying vorticity $L=\{0,1\}$ can be realized in the bridge.

\acknowledgements

We thank S.~Vieira and H.~Suderow for their remarks and for critical reading
of the manuscript in its preparation phase. We acknowledge discussions with
V.V.~Moshchalkov, L.~Van~Look and V.~Bruyndoncx. This work has been
supported by the IUAP, the F.W.O.-V. projects Nos. G.0287.95, 9.0193.97,
G.0306.00, G.0274.01N, W.O.G. WO.025.99N, GOA BOF UA 2000 (Belgium), and the ESF
Programme VORTEX.

\newpage

\noindent {\bf Figure captions}

\bigskip

\noindent Fig.~1. Scheme of the model of the mesoscopic superconducting
bridge. The bridge consists of six conoids symmetric with respect to the
plane $z=0$. Sizes of the model bridge used in our calculations: $b_{0}=2.6$
nm, $b_{1}=8.9$ nm, $b_{2}=15.8$ nm, $b_{3}=23.0$ nm, $h_{1}=3.6$ nm, $%
h_{2}=11.7$ nm, $h_{3}=15.8$ nm. In the insert, we show a schematic plot 
of the bridge used in the experiment \cite{poza}. 

\bigskip

\noindent Fig.~2. Distribution of the superconducting phase, $\left| \psi
(x,y,z)\right| ^{2}$, as a function of applied magnetic field, in the
nanosized superconducting bridge (a contour plot is shown for the cross-section 
$y=0$) for $L=0$, $T/T_{c}=0.6$. 
Sizes are shown in nanometers for the Pb bridge described in Ref.~\onlinecite{poza}. 

\bigskip

\noindent Fig.~3. Distribution of the superconducting phase, $\left| \psi
(x,y,z)\right| ^{2}$, as a function of temperature, in the nanosized
superconducting bridge (a contour plot is shown for the cross-section $y=0$) 
for $L=0$, $H_{0}=5H_{c2}$. 

\bigskip

\noindent Fig.~4. Distribution of the superconducting phase, $\left| \psi
(x,y,z)\right| ^{2}$, as a function of the orbital quantum number $L$, in
the nanosized superconducting bridge (a contour plot is shown for the cross-section 
$y=0$) for $T/T_{c}=0.6$, $H_{0}=5H_{c2}$. 

\bigskip

\noindent Fig.~5. Free energy $F_{s}-F_{n}$ [measured in $H_{c}^{2}(0)/4\pi $] 
of the nanosized superconducting bridge, as a function of the applied 
magnetic field, for $T/T_{c}=0.4$ and various $L$. 

\bigskip

\noindent Fig.~6. Free energy $F_{s}-F_{n}$ [measured in $H_{c}^{2}(0)/4\pi $%
] of the nanosized superconducting bridge, as a function of the applied
magnetic field, for $T/T_{c}=0.6$ and various $L$.

\bigskip

\noindent Fig.~7. Free energy $F_{s}-F_{n}$ [measured in $H_{c}^{2}(0)/4\pi $%
] of the nanosized superconducting bridge, as a function of the applied
magnetic field, for $T/T_{c}=0.9$ and various $L$.

\bigskip

\noindent Fig.~8. Resistance of the superconducting bridge as a function of
the applied magnetic field for $L=0$ and various temperatures: $T/T_{c}=0.4$
(a), $T/T_{c}=0.6$ (b), and $T/T_{c}=0.9$ (c). For the case $T/T_{c}=0.6$,
the states with $L\neq 0$ are also plotted.

\bigskip

\noindent Fig.~9. Free energy $F_{s}-F_{n}$ [measured in $H_{c}^{2}(0)/4\pi 
$] as a function of the applied magnetic field, for the mesoscopic disks
with diameters coinciding with the diameters of the bridge in various
cross-sections normal to the $z$-axis. The free energy is calculated for for 
$T/T_{c}=0.6$ and various $L$. The diameters of the disks are: $D=1.5\xi$ [24 nm 
for the Pb bridge\cite{poza}] (a), $D=1.25\xi$ (20 nm) (b), 
$D=1.125\xi$ (18 nm) (c), $D=0.75\xi$ (12 nm) (d), $D=0.5\xi$ (8 nm) (e). 

\bigskip

\noindent Fig.~10. Distribution of the superconducting phase, $\left| \psi
(x,y,z)\right| ^{2}$, as a function of the orbital quantum number $L$, in
the nanosized superconducting bridge (a contour plot is shown for the cross-section 
$y=0$) for $T/T_{c}=0.6$, $H_{0}=5H_{c2}$: $L=0$ (a), $L=1$ (b), 
and the state characterized by a variable vorticity $L=\{0,1\}$ (c). 


\begin{references}
\bibitem{uncert} It is worth noting, that some uncertainty in this parameter
does not lead to any substantial modifications of our results. Indeed, $\xi $
enters $\kappa =\lambda /\xi $, which is the only parameter of the GL
theory. Our analysis shows, that relative changes of  $\left| \psi \right|
^{2}$ in any point of the bridge are at most 10 percent when increasing $%
\kappa $ by a factor of 2.
\end{references}
\end{document}